# New Abeans for TINE Java Control Applications

J. Dovc, M. Kadunc, Y.Oku, M. Plesko, G.Tkacik J. Stefan Institute
K. Bartkiewicz and P. Duval, DESY

Abstract

"Abeans" [1] (wide-interface accelerator Java beans developed at JSI) have in the past been used with great success in control applications at ANKA and ESO and on test cases at the SLS and Riken. At DESY, "TINE" [2] is used as the principal control system for HERA as well as the intercommunication protocol among the HERA experiments. To date, most TINE-based client-side applications have been written using ACOP [3] (a narrow-interface accelerator component) in Visual Basic (VB), which has provided a remarkably powerful developing environment for generating professional control applications. Currently, however, VB control applications can only run on Windows-based desktop machines and consoles. As it is often desirable to provide certain control applications on non-Windows platforms (or indeed over the Web), we have created an equally powerful developing environment based on Java and the next release of Abeans, where an ACOP-like narrow-interface bean has been developed. Like ACOP, the Abean itself accepts plugs from various communication protocols, but was brought to fruition at DESY using the TINE Java class. Details concerning matching Abeans and TINE, as well as the pros and cons of "wide" versus "narrow" interfaces will be presented below. Several applications will also presented along with the results of benchmarking against similar VB applications.

## 1 INTRODUCTION

For some time the mantra of RAD-aficionados has been "components." A component is a reusable object, dedicated to a specific task, offering a well-defined interface consisting of properties, methods, and events, and which can be plugged into an application during the design phase. In the Windows world these are typically ActiveX controls and in the Java world, Java beans.

There are two components targeting accelerator control which we shall compare and contrast in what follows, namely Abeans[1] and ACOP[3]. In particular, as the latest release of Abeans features a plug for the TINE [2] control system as does ACOP, we can compare the same application using both ACOP and Abeans.

## 2 ABEANS RELEASE 3

Abeans R3 is a library that provides simple Java beans for connection with the control system. At the same time, it provides several useful services: logging, exception handling, configuration and data resource loaders, authentication, and policy management.

As Abeans are designed to run multiple applications in a single JVM (Java Virtual Machine), libraries are loaded only once and the memory footprint is modest compared to applications running in separate JVMs. This feature also allows the same applications to be run individually or from within an applet in a web browser.

In Abeans, different *models* are used to represent the structure of the control system. Models use *plugs* to get data from a specific control system. At DESY we used the Abeans "channel" model (i.e. a narrow interface access model), which consists of namespaces and channels, to create a plug connecting the TINE Java class to the Abeans.

Like Abeans, ACOP accepts plugs to any control system communication layer. At DESY, TINE is most widely used. It was therefore natural to rewrite several controls applications at DESY using Abeans, offering a chance to contrast their differences.

## 3 ABEANS CHANNEL AND TINE

An Abeans *Namespace* object is a container of elements and its functionality is limited to retrieving names and types of its children (which can be either namespace or channel). In the TINE control system namespaces represent contexts, servers and devices. A *Channel* represents a single property in the control system, specified by its name, plug type and data type (single-value or array). Several Channel classes were implemented to map the TINE data types to standard Java types, e.g. TINE types tdouble and tfloat map to Abeans' DoubleChannel bean. The Channel bean provides synchronous read-write and asynchronous read-only access to TINE properties, access to their history and miscellaneous data such as minimum and maximum value, engineering units, description, type of access (read or write) and size for array properties. The data are retrieved simply by a method call to the Channel object and are converted to a standard Java type. For example:

```
BasicRemoteInfo info = new
BasicRemoteInfo("/HERA/HEPBPM/WL197_MX/ORBIT.X","Channel-TINE");
DoubleChannel channel=new DoubleChannel(defaultFamily, info);

double      singleValue    =    channel.getValue(new Completion());
double[]    multipleValues =    channel.getValue(141,new Completion());
double      minimum        =    channel.getMinimum();
String      units          =    channel.getUnits();
```

The Channel bean significantly reduces the amount of code needed to retrieve data from the control system, handling all initialization, termination, data acquisition and conversion, and thread management, as well as hiding the communication protocol from the application programmer.

Another significant advantage of using Abeans and Channels are plugs. The same application can use any plug that supports the channel model, so it is possible to create a simulator plug to test the applications without actual access to the servers.

## 4 WIDE VS. NARROW INTERFACE

Abeans were originally designed for wide-interface control system at ANKA. At DESY, a narrow-interface to the TINE protocol was implemented using a new release of Abeans where the "channel" model was applied. We now contrast these approaches.

A wide interface is very object-oriented and easy to use. Here each device is represented by an object whose properties and methods reflect the state and functionality of the device. A physicist can then easily imagine the devices and their properties and access them in an intuitive manner. There is one data source for all the data connected with a specific object. Note however, that each property must be well defined, requiring a long design phase. The payoff comes with maintenance, documentation and general ease in programming. On the other hand, adding functionality can be quite complicated and time consuming since a change in the definition of a device requires a reflected change in the application interface (otherwise the pieces of the control system loose their compatibility). Fortunately, the change will be intrinsic to the data itself and not break the user interface.

Applications which profit the most from this approach are those which are focused on a small number of devices and access several properties and parameters of the same device rather than a single property of multiple devices. In the latter case many objects would be created when only a fragment of each object's functionality is needed. In such cases the narrow interface approach is more appropriate.

Narrow-interface access consists of a single family of entities (properties), which are identified by their name (which can be hierarchical). Properties have a default representation, and provide other descriptive data (type, bounds, description, units etc). Properties from a server are often connected at the device level, and one property can return values from all the devices on the server. This is very useful when one knows the structure of the properties on a given server. Data requires less processing and network bandwidth and is more easily applied to different graphical displayers. Applications generally run faster.

However, such a system is much more difficult to maintain. Once a server is developed, there is frequently little or no written documentation on the organization of the properties. As a client-side application programmer using the system must be acquainted with the structure of the server he is accessing, he is sometimes put in the position of seeking out the server-developer for detailed information. Thus, if a narrow interface access is used, it should be well organized and consistency should be assured in order to help maintain the system and reduce complexity in application development.

## 5 DESY APPLICATIONS

Some core DESY Visual Basic applications were rewritten in Java in order to make them available on other operating systems and to benchmark them against similar VB applications. Those applications are Instant Client, Vacuum Panel, BPM Panel and Archive Reader.

Applications can be run as both applets in a web browser or individually. An applet is a program written in the Java programming language that can be included in an HTML page, in a similar way an image is included. When you use a Java technology-enabled browser to view a page that contains an applet, the applet's code is transferred to your system and executed by the browser's Java Virtual Machine (JVM) or Java plug-in, depending on html tag, used in html page. Because browser's JVM supports only Java 1.1, a Java plug-in is used (Java plug-in Home Page: http://java.sun.com/products/plugin/index.html). A special launch application, developed as an applet, takes care of launching particular programs. Because of the Abeans design they can run inside the same JVM. Because all XML configuration files and jar libraries can be stored on a web server, a client requires only

JRE 1.2 or later and internet browser that support Java plug-in. Applications can be run without any special installation procedure.

We are now ready to report on the first few TINE Abeans applications. These will include the TINE "Instant Client", the TINE "Archive Reader", and a vacuum display panel. The applications presented here were developed using Visual Age for Java, thus a comparison of application development using Visual Basic will follow.

## 6 RESULTS

In reporting results, we will draw a direct comparison of the applications mentioned above, generated in Visual Age for Java (VAJ), and their counterparts generated in Visual Basic (VB).

Let us first consider the development environment. When one is at home in his development environment, this sometime boils down to simply a matter of taste. Qualitatively, we can only say that the learning curve for VAJ is much steeper than for VB. The debug cycle is also more lengthy for VAJ, as expected since VB is only single-threaded. The VB integrated developing environment (IDE) also loads much faster than that for VAJ, and requires much fewer local resources (memory). The VB IDE typically loads in less than 10 seconds (even when loading a large project) from scratch and takes up around 10 Mbytes of main memory. VAJ requires somewhat more time (~25 seconds) and ends up consuming close to 40 Mbytes of main memory. The user must also develop a different intuition for visual editing in VAJ than for VB. Basically however, these are all relatively minor points.

Generally speaking, native-compiled VB programs start more quickly than Java programs, particularly if the JVM must also be loaded. However, the effect is smaller than one might think. For instance the VB Vacuum Panel loads and begins running in approximately 2 seconds if started from the local disk. The VAJ Vacuum Panel loads in perhaps 5 seconds under similar circumstances (as an application started from the local disk). Once up and running, however, there is no perceivable (or measurable for that matter) difference concerning data acquisition over the TINE interface, approximately 2 milliseconds per synchronous call.

Where the VAJ applications suffer in comparison with VB is the case where multiple control applications are run on the same desktop. As one might expect, native applications end up using far fewer system resources (mostly memory) than Java applications. Where there is no problem in running 20 or more VB control applications on a desktop (the only limitation is the physical memory), the operating system will begin to struggle with as few as 4 Java applications. However, the design of Abeans R3 significantly improves this situation, as it allows running several independent applications within a single virtual machine Of course, there is no way to run even 1 VB application on a Linux machine!

## 7 CONCLUSIONS

The results of the preceding section suggest that Java is slightly off the mark set by VB as concerning ease of development and performance. However, the object oriented (OO) features of Java are much better than for VB (ignoring for the moment VB.NET), and where this is important, Java provides a much better programming environment. There is also a tremendous advantage of the platform independence offered by Java. The question remains as to just how important a full set of OO features and platform independence really are for console applications.

There is as yet no rush from VB to Java, nor is there a reason to do so. However the addition of Abeans plus TINE at DESY has been enthusiastically welcomed, and those persons who prefer Java now have an accelerator toolkit. Although accelerator control proper at DESY uses VB programs, a growing number of browser launched diagnostic programs (using Abeans) are being generated and made available. This has enriched both the Abeans portfolio and the TINE portfolio considerably.